\documentclass[11pt]{article}
\usepackage{graphicx} 

\usepackage{fancyhdr}
\usepackage{isomath}
\usepackage{mathtools} 
\usepackage{amsbsy}
\usepackage{amssymb}
\usepackage{amscd,amsfonts}
\usepackage{bigints}
\usepackage{graphicx}
\usepackage{verbatim}
\usepackage{euscript}
\usepackage{alltt}
\usepackage{stmaryrd}
\usepackage{relsize}
\usepackage{enumerate}
\usepackage{url}
\usepackage[stable]{footmisc}
\usepackage{hyperref}
\usepackage{breakurl}
\usepackage{comment}

\usepackage[maxbibnames=9,sorting=none,style=numeric]{biblatex}
\usepackage[font=small,labelfont=bf]{caption}
\usepackage[font=small,labelfont=bf]{subcaption}
\usepackage{float}
\usepackage{mwe}
\usepackage{algorithm}
\usepackage{algpseudocode}
\usepackage{xcolor}
\usepackage[super]{nth}
\usepackage{soul}

\DeclareGraphicsExtensions{.eps,.pdf}
\usepackage{amsmath}
\usepackage{amsbsy}
\usepackage{amssymb}
\usepackage{amscd}
\usepackage{amsfonts}
\usepackage{isomath}

\newcommand{\beq}{\begin{equation}}
\newcommand{\eeq}{\end{equation}}
\newcommand{\beqs}{\begin{eqnarray}}
\newcommand{\eeqs}{\end{eqnarray}}
\newcommand{\beql}{\begin{equation} \label}
\newcommand{\half}{\frac{1}{2}}

\newcommand{\calJ}{{\cal J}}

\newcommand{\calO}{{\cal O}}

\newcommand{\p}{\partial}
\newcommand{\dee}{\mathcal{D}}
\newcommand{\scl}{\mathcal{L}}

\DeclareMathOperator*{\argmax}{argmax}

\usepackage[margin=1in]{geometry}

\date{}
\begin{document}
\title{The Second Law as a constraint and admitting the approximate nature of constitutive assumptions}

\author{Amit Acharya\thanks{Department of Civil \& Environmental Engineering, and Center for Nonlinear Analysis, Carnegie Mellon University, Pittsburgh, PA 15213, email: acharyaamit@cmu.edu.}}

\maketitle
\begin{abstract}
\noindent A scheme for treating the Second Law of thermodynamics as a constraint and accounting for the approximate nature of constitutive assumptions in continuum thermomechanics is discussed. An unconstrained, concave, variational principle is designed for solving the resulting mathematical problem. Cases when the Second Law becomes an over-constraint on the mechanical model, as well as when it serves as a necessary constraint, are discussed.

\end{abstract}

\section{Introduction}
Ever since the pioneering work of Coleman and Noll \cite{CN} utilizing the local statement of the Second Law of Thermodynamics in the form of the Clausius-Duhem Inequality (Truesdell and Toupin \cite[Sec.~258, discussion surrounding (258.4)]{tr-to}), the Second Law has been used as a constraint on admissible constitutive equations in continuum thermomechanics. As discussed subsequently in this Section,  beyond sound theory, there are good practical reasons to do so, related to a question of constraints. However, notions of temperature and entropy are physically elusive, and it is perhaps fair to say that the most direct physical grasp of these concepts comes only from their association with ideas from statistical mechanics, as described, e.g., by Berdichevsky \cite{berd} (a recipe for the combination of these concepts with continuum thermomechanics is explored in \cite{acharya2011microcanonical} in the context of dislocation thermomechanics). When/if viewed from this perspective, it becomes amply clear that the time-scales involved in posing the notion of entropy cannot be arbitrarily small, and entropy is not a quantity that immediately lends itself to a definition of an entropy density per unit mass, as is assumed in continuum thermomechanics. Indeed, in recent work Needleman \cite{needleman2024perspective, needleman2023discrete} has questioned the applicability of  the Second Law in the form of the Clausius-Duhem Inequality \cite{tr-to} over small time and length scales. It is known that the Second Law, in general, is not a guarantor of `stability' of thermomechancial processes and it is also shown in \cite{needleman2023discrete} that a choice of constitutive assumptions in a simple example of wave propagation in a viscoplastic material which allows for the violation of the Second Law over small periods of time and over small length scales does not necessarily lead to some irreversible, catastrophic loss of stability of motion. Needleman's conclusion in \cite{needleman2023discrete} is that the unequivocal satisfaction of the Clausius-Duhem Inequality almost everywhere in space-time during the course of every process of a body/system is too strong a requirement for small-scale phenomena.

In this work, an alternate, possibly complementary, point of view is explored through a relatively simple example whose ideas carry over to general processes of materials described by local constitutive assumptions. `Excess' quantities in typically constitutively specified fields, e.g., stress, are introduced to acknowledge the inevitable lack of full knowledge in describing the material response of any body that is modeled. Then the (mechanical form of the) Clausius-Duhem Inequality (alternatively referred to as the Second Law in this work) is assumed to hold, simply as a constraint condition on physical processes, much like the balances of mass, linear and angular momentum, and energy. It is required that the excess quantities be `as small as possible.' This gives rise to a constrained nonlinear optimization problem. A procedure is then outlined to turn the problem into an unconstrained concave variational principle with guaranteed consistency with the PDE describing the mechanical model and the Second Law, including equal number of fields as the number of equations. Contact is made with Needleman's  point of view \cite{needleman2024perspective, needleman2023discrete} by noting that if the Clausius Duhem Inequality is satisfied with only the excess stored energy in play with all other mechanical constitutive assumptions unaltered, then removal of the former from a given solution can very well lead to the violation of the Second law over some regions of the body and over certain periods of time for the process. This is discussed further in Sec.~\ref{sec:concl}.

This brief communication contains three more sections: in Sec.~\ref{sec:fix} we fix ideas in the context of a nominally elastic, 1-d bar undergoing a quasi-static deformation history. In Sec.~\ref{sec:dual} the corresponding elastodynamic problem is defined, a specific recipe for solving the problem through a variational principle is developed, and a concave variant elaborated on. Sec.~\ref{sec:concl} contains some concluding remarks.

In terms of notation, subscript $x,t$ will represent partial derivatives w.r.t the quantities. A superscript prime on a function of one variable represents a derivative. `Arbitrarily' specified fields are understood to have the requisite regularity for the mathematical statements they appear in to make sense.

\section{Fixing the physical idea}\label{sec:fix}
To fix ideas, consider quasistatic elasticity in 1-d along with a mechanical version of the Second Law (cf.~\cite[Sec.~3]{needleman2024perspective}):
\begin{equation}\label{eq:qs_elast}
    \begin{aligned}
        T_x & = 0 \\
        T e_t - \psi_t &\geq 0\\
        v(0,t) = 0; &\qquad T(1,t) = f(t),
    \end{aligned}
\end{equation}
where $T = \hat{T}(e)$ is the stress, $e$ is the strain, and $\psi = F(e)$ is the (free) energy density, $\hat{T}$ and $F$ are arbitrarily specified constitutive response functions, and $t \mapsto f(t)$ is an applied load history.  Then, force equilibrium  $(T(x,t) = f(t))$ determines
the strain trajectory, say $(x,t) \mapsto e^m(x,t)$ given some initial condition on $e$, which satisfies
\begin{equation*}
    e^m_t(x,t) = \frac{f_t(t)}{\hat{T}'(e^m(x,t))}.
\end{equation*}
The Second Law requires
\[
\left( \hat{T}(e^m(x,t)) - F'(e^m(x,t)) \right) e^m_t(x,t) \geq 0,
\]
and for specified $\hat{T}, F$ there is no reason why such a requirement would be satisfied by the generated strain trajectory $e^m$. One (and perhaps the only currently known) option out of this quandary is to use the Coleman-Noll \cite{CN} procedure to restrict the constitutive response functions which, in this case, would amount to requiring
\[
\hat{T}(e) = F'(e) \mbox{ for all } e.
\]

We would now like to explore a different tack on the above question. Based on the premise that the modeler is in a position to deploy physically sound constitutive assumptions (see, e.g., the discussions in \cite{needleman2023discrete,needleman2024perspective}) but at the same time admitting that there is no natural way to know what an exact constitutive equation for a material under study should be, it would be ideal to assume that the stress and free energy density for a nominally elastic material are given by the specifications
\begin{equation}\label{eq:const_ex}
\begin{aligned}
        T(x,t) & = \hat{T}(e(x,t)) + X(x,t)\\
    \psi(x,t) & = F(e(x,t)) + a(x,t)
\end{aligned}
\end{equation}
where, as before, $\hat{T}, F$ are specified approximate constitutive equations, and $X,a$ are excess fields to compensate for the lack of exact knowledge of the material, to be determined by the overall problem to be solved under the constraint that satisfying the balance principles \eqref{eq:qs_elast} is non-negotiable.

Operating with these design constraints for solving \eqref{eq:qs_elast}, we have that the trajectories
\[
(x,t) \mapsto (e(x,t), X(x,t), a(x,t))
\]
must satisfy 
\[
\hat{T}(e(x,t)) + X(x,t) = f(t), \qquad \psi_t(x,t) = F'(e(x,t))\,e_t(x,t) + a_t(x,t)
\]
and (dropping the $(x,t)$ argument)
\[
\big( \hat{T}(e) - F'(e) \big) e_t + X e_t - a_t = \big( f - F'(e) \big) e_t - a_t \geq 0.
\]
One is now faced with the opposite problem of the situation being not sufficient to determine a unique strain trajectory, in general. Simply consider the situation $a = 0$ for all times and then $e(x,t) = 0$ for all $(x,t)$ is a solution as well as one that solves $F'(e(x,t)) = f(t)$ for given $F$ and, for simplicity, consider the linear elastic situation when $F' = Ce$, $C$ the constant Young's modulus, and one subsequently determines $X(x,t) := f(t) - \hat{T}(e(x,t))$. In more than 1 space dimension, $X$ would be a tensor and the Second Law, a scalar inequality, would be even less constraining.

With the desire to retain the notion of excess fields in the constitutive structure as in \eqref{eq:const_ex}, we augment it by introducing a scalar, real-valued \emph{dissipation slack} variable, $s$ (with $s^2$ the physical \emph{dissipation}), and require that \eqref{eq:qs_elast} be satisfied in the form
\begin{equation*}
    \begin{aligned}
        T_x & = 0 \\
        T e_t - \psi_t - s^2 &= 0\\
        v(0,t) = 0; &\qquad T(1,t) = f(t),
    \end{aligned}
\end{equation*}
augmented with the further conditions that
\[
|X(x,t)|^2, \quad |a(x,t)|^2, \quad - |s(x,t)|^2 \quad \mbox{ for all } (x,t) \mbox{ in the space-time domain under consideration}
\]
\emph{be minimized} in any admissible process of the material under consideration; the minus sign on the dissipation is to account for the nature of the dissipation which one wants to maximize. To get a rough sense of what this may result in, assuming $X = 0$, $a = 0$ one would require
\begin{subequations}
    \begin{align}
        \hat{T}(e(x,t)) &= f(t) \label{eq:ex-fix1}\\
        \big( \hat{T}(e(x,t)) - F'(e(x,t)) \big) e_t (x,t) &= 0
    \end{align}
\end{subequations}
for specified functions $\hat{T}, F$ and if a strain trajectory cannot be found satisfying the above, $X= 0, a = 0$ would not be attained and the constrained minimization problem would be required to determine the solution. For example, assuming the constraint eqn. \eqref{eq:ex-fix1} determines a strain trajectory with $|X|$ allowed to take its minimum value, $a_t$ will have to compensate for $(f - F'(e))e_t$ when the latter is negative, and when positive, $s^2$ will have to compensate for it with $a_t$ allowed to vanish - this, assuming that the cost of deviations in $|X|, |a|$ from $0$ is large in comparison to that of $|s|$, and that $a$ belongs to a function space where if $a$ is small in any interval of time, its time derivative in the interval cannot be large.

Of course, the requirement of solving a nonlinear constrained optimization problem is not a recipe for making life easier in continuum mechanics, but a scheme for converting such problems to an unconstrained \emph{concave} variational problem can provide some advantages and, at the very least, a different point of view. It is such a scheme that we develop in Section \ref{sec:dual}.

In closing this Section, we emphasize that the device of allowing excess fields is not a declaration of more sophisticated knowledge about material response, but rather an admission of the lack of such knowledge. Thus, what we propose is not meant as a substitute for the development of more physical and fundamental models of the response of materials (e.g., \cite{chatterjee2020plasticity,acharya2011microcanonical,acharya2023dual}). It is well-understood that even in the absence of the excess fields and adopting standard constitutive equations, problems for specific material models (e.g.~nonconvex elastodynamics discussed in Sec.~\ref{sec:concl}, softening geomaterial response) often display non-uniqueness of solutions. The proposed scheme below has the potential of exploring such non-unique solutions with a modicum of stability in problem solving \cite{singh2024hidden} to check whether any of those can correspond to actual material response, without having to rely on ad-hoc higher-order regularizations or extra constitutive statements introduced solely for the purpose of mathematical well-posedness. 

\section{Dual Elastodynamics with excess fields}\label{sec:dual}
We consider the stress and free-energy response of a nominally elastic material with constitutive response not known with definiteness as given by \eqref{eq:const_ex}, where $X,a$ are constitutively undetermined fields required to be determined by the model. We then write the equations of such a model of elastodynamics comprising kinematic compatibility between the velocity and the strain fields, balance of linear momentum (without body force for simplicity), and the Second Law of thermodynamics as 
\begin{equation}\label{eq:primal_basic}
    \begin{aligned}
        v_x - e_t & = 0 \\
        r v_t - \hat{T}(e)_x - X_x & = 0 \\
        \left(\hat{T}(e) + X - F'(e) \right) e_t - a_t & \geq 0,
    \end{aligned}
\end{equation}
where $v$ is the velocity, $e$ the strain, and $r$ the constant mass density. We denote $\hat{T}(e) - F'(e) =: T^\sharp(e)$, and write \eqref{eq:primal_basic} in first-order form with the requirement that no nonlinearities involving derivatives be present in any of the equations, excluding terms that may be expressed as space or time derivatives of nonlinear functions of the basic dependent variables, e.g., $\hat{T}(e)_x$:
\begin{subequations}\label{eq:primal}
\allowdisplaybreaks
\begin{align}
     v_x - e_t & = 0 \notag\\
     r v_t - \hat{T}(e)_x - X_x & = 0 \notag\\
     e_t - d & = 0 \tag{\ref{eq:primal}}\\
     \left(T^\sharp(e) + X \right) d - a_t - s^2 & = 0 \notag\\
     s_x - g & = 0, \notag
     \end{align}
\end{subequations}
where $s$ is the dissipation slack, a real-valued function of space-time to be determined by the model.
For simplicity, but not out of necessity, we will assume
\[
T^\sharp(e) = 0; \qquad a = 0
\]
 in this Section (which would be an exact statement and not an assumption if the material model considered was hyperelastic). Also, the last equation in \eqref{eq:primal} at this point assigns a name to the spatial gradient of the dissipation slack, whose purpose will be to impose a regularization in the spatial gradients of the dissipation when deemed desirable by the modeler, further discussed in Sec.~\ref{sec:concl}. We augment the system with the boundary and initial conditions, say,
\begin{equation}\label{eq:ibc}
        v(0,t) = v^l(t); \qquad T(1,t) = f(t); \qquad e(x,0) = e^0(x); \qquad v(x,0) = v^0(x).
\end{equation}

The goal now is to define an optimization problem whose solution satisfies the system (\ref{eq:primal}-\ref{eq:ibc}) with the fields $(X,s,g)$ point-wise being as small as possible so that the solution is as close as possible to an admissible process of a hyperelastic material. 

To do so following \cite{ach_HCC}, define a \emph{pre-dual} functional, $\widehat{S}[U,D]$, of primal fields
\[
U := (v,e,X,d,s,g)
\]
and dual fields
\[
D := (\lambda, \mu, \beta, \rho, \gamma), \qquad \dee := (D, D_x, D_t)
\]
given by,
\begin{equation}\label{eq:pre-dual}
\begin{aligned}
     \widehat{S}[U,D] & := \int_0^{t_f} \int_0^L \scl(U,\dee, \bar{U}) \, dx dt \\
  & \quad - \int_0^{t_f} v^l(t) \lambda(0,t) \, dt + \int_0^L e^0(x) \lambda(x,0) \, dx - \int_0^L r v^0(x) \mu(x,0) \, dx - \int_0^{t_f} f(t) \mu(1,t) \, dt,\\
  \mbox{ where }& \\
   \quad \scl(U,\dee) & := - v \lambda_x + e\lambda_t - r v \mu_t + \left( \hat{T}(e) + X \right)\mu_x + \rho X d - \rho s^2 - s \gamma_x - \gamma g  - e \beta_t - \beta d +  H(U, \bar{U}), \\
   \quad H(U, \bar{U})  & := \half \left( c_X X^2 + c_s s^2 + c_g g^2 \right) + \half c_d \left(d - \bar{d} \right)^2 + \half c_e \left( e - \bar{e} \right)^2 + \half c_v \left( v - \bar{v} \right)^2,\\
   \quad \bar{U} & :=  (\bar{v},\bar{e},0,\bar{d},0,0).
\end{aligned}
\end{equation}
In the above, $c_X, c_g, c_d, c_e, c_v$ are arbitrarily chosen large real-valued positive constants, $c_s$ is a negative $O(1)$ constant, and $(x,t) \mapsto (\bar{d}(x,t), \bar{e}(x,t), \bar{v}(x,t))$ are arbitrarily specifiable functions of space-time. The constants are required to have physical dimensions so that all additive  terms in $H$ have the same physical dimensions, of course. Thus, we maximize the dissipation, subject to the mechanical constraint equations.

We also impose Dirichlet boundary conditions on the functions
\begin{equation*}
    \begin{aligned}
        \lambda(L, t) = \lambda^l(t), \quad \lambda(x,t_f) = \lambda^t(x), \quad \mu(x,t_f) = \mu^t(x),\\
        \mu(0,t) = \mu^l(t), \quad \beta(x,0) = \beta^b(x), \quad \beta(x,t_f) = \beta^t(x),
    \end{aligned}
\end{equation*}
where the functions on the right of the equality signs are arbitrarily specified.

The pre-dual functional may be viewed as one generated from treating the system \eqref{eq:primal} as constraints with a view towards minimizing the objective $H$ by the method of Lagrange multipliers, the dual fields.

For illustrative purposes, we work with a constitutively-determined stress response function of the form
\begin{equation}\label{eq:stress_const}
    \hat{T}(e) := E_l e + E_* e_*^2 - E_* (e - e_*)^2
\end{equation}
for $E_l, E_*$ positive elastic moduli and $e_* > 0$ is a threshold strain, such that $E_l + 2 E_* e_* = E$, where $E$ is the Young's modulus of the material. Since
\[
\hat{T}'(e) = E_l - 2 E_*(e - e_*),
\]
the model shows softening for strains in the range
\[
\frac{E_l}{2 E_*} + e_* < e.
\]
In the arbitrary choice of the parameter $c_e$ in the definition of the potential $H$, we impose a lower limit of $c_e > 2 E_*$.

Next, we will consider the first-order optimality conditions for the optimization of $\widehat{S}$ w.r.t $(U,D)$ in slightly non-standard manner. Instead of looking for a pair $(\hat{U}, \hat{D})$ that satisfies
\begin{equation*}
\begin{aligned}
     0 & = \delta \widehat{S}[\hat{U},\hat{D}; \delta U, \delta D] \\
     & = \int_0^{t_f} \int_0^L \left( \p_U \scl(\hat{U},\hat{\dee} , \bar{U})  \delta U + \p_\dee\scl(\hat{U},\hat{\dee}, \bar{U}) \delta \dee \right) \, dx dt \\
  & \quad - \int_0^{t_f} v^l(t) \delta \lambda(0,t) \, dt + \int_0^L e^0(x) \delta \lambda(x,0) \, dx - \int_0^L r v^0(x) \delta \mu(x,0) \, dx - \int_0^{t_f} f(t) \delta \mu(1,t) \, dt
\end{aligned}
\end{equation*}
for all $(\delta U, \delta D)$ constrained by
\begin{equation}\label{eq:dual_dirichlet_var}
    \begin{aligned}
        \delta \lambda(L, t) = 0, \quad \delta \lambda(x,t_f) = 0, \quad \delta \mu(x,t_f) = 0,\\
        \delta \mu(0,t) = 0, \quad \delta \beta(x,0) = 0, \quad \delta \beta(x,t_f) = 0,
    \end{aligned}
\end{equation}
we first look for a \emph{dual-to-primal} DtP mapping
\[
U = U_H(\dee, \bar{U})
\]
such that
\begin{equation}\label{eq:dtp_cond}
\p_U \scl \left(U_H(\dee, \bar{U}),\dee, \bar{U} \right) = 0
\end{equation}
is satisfied for all $(\dee, \bar{U}) \subset \calO$, the latter being some open subset of the (finite dimensional) space in which $(\dee, \bar{U})$ take values. Then we define a \emph{dual functional}
\begin{equation}\label{eq:dual}
\begin{aligned}
     S[D] & := \int_0^{t_f} \int_0^L \scl \left(U_H(\dee, \bar{U}),\dee, \bar{U} \right) \, dx dt \\
  & \quad - \int_0^{t_f} v^l(t) \lambda(0,t) \, dt + \int_0^L e^0(x) \lambda(x,0) \, dx - \int_0^L r v^0(x) \mu(x,0) \, dx - \int_0^{t_f} f(t) \mu(1,t) \, dt,
\end{aligned}
\end{equation}
and look for a $\hat{D}$ such that $(\hat{\dee} (x,t), \bar{U}(x,t)) \in \calO$ for $(x,t)$ a.e.~in $(0,L) \times (0,t_f)$ and 
\begin{equation}\label{eq:dual_first_var}
\begin{aligned}
     0 & = \delta S[\hat{D}, \bar{U}; \delta D] \\
     & = \int_0^{t_f} \int_0^L  \p_\dee \scl \left(U_H(\hat{\dee}, \bar{U}), \hat{\dee}, \bar{U} \right) \delta \dee \ \, dx dt \\
  & \quad - \int_0^{t_f} v^l(t) \delta \lambda(0,t) \, dt + \int_0^L e^0(x) \delta \lambda(x,0) \, dx - \int_0^L r v^0(x) \delta \mu(x,0) \, dx - \int_0^{t_f} f(t) \delta \mu(1,t) \, dt
\end{aligned}
\end{equation}
for all $\delta D$ constrained by \eqref{eq:dual_dirichlet_var}. Noting that $\scl$ is necessarily affine in $\dee$ with the coefficient of $\dee$ arising from integration by parts of the primal system \eqref{eq:primal}, it can be seen that \emph{the Euler-Lagrange equations and side conditions of the dual functional $S$ are exactly the primal system \eqref{eq:primal} and the conditions \eqref{eq:ibc} with the substitution $U \to U_H$.}

We note that \emph{the number of equations in the Euler-Lagrange system of $S$ equals the number of dual fields (here 5) which also equals the number of primal equations \eqref{eq:primal}, regardless of the number of primal fields involved (here 6).} The solution for the 6 primal fields are obtained from the DtP mapping after solving for the 5 dual fields.

Another important point to note is that as long as a DtP mapping satisfying \eqref{eq:dtp_cond} can be constructed by a choice of an auxiliary potential $H$, the use of any such $H$ guarantees that the E-L equations of the corresponding dual functional is identical to the primal system with the replacement $U \to U_H$.

We now implement the above strategy for the specific constitutive equation \eqref{eq:stress_const} for stress and the choice of the auxiliary potential $H$ displayed in \eqref{eq:pre-dual}. The DtP mapping is given by
\begin{subequations}\label{eq:dtp}
    \allowdisplaybreaks
    \begin{align}
        \p_v \scl = 0: \qquad & c_v \left( v_H - \bar{v} \right) = \lambda_x + r \mu_t \notag\\
        \p_e \scl = 0: \qquad & (c_e - 2 E_* \mu_x) \left( e_H - \bar{e} \right) = - \lambda_t - E_l \mu_x + 2 E_* \mu_x (\bar{e} - e_*) + \beta_t  \notag\\
        \begin{bmatrix}
            \p_d \scl \\
            \p_X \scl
        \end{bmatrix} = 0: \qquad & \begin{bmatrix}
             c_d & \rho \\
             \rho & c_X
         \end{bmatrix} \begin{bmatrix}
                         d_H- \bar{d} \  \\
                          X_H
                        \end{bmatrix} = \begin{bmatrix}
                                           \beta \\
                                           - \mu_x - \rho \bar{d} \ 
                                        \end{bmatrix} ; \qquad \mathbb{K}(\dee) := \begin{bmatrix}
                                                                                      c_d & \rho \\
                                                                                       \rho & c_X
                                                                                    \end{bmatrix}\tag{\ref{eq:dtp}} \\
          \p_s \scl = 0:  \qquad & (c_s - 2 \rho) s_H = \gamma_x \notag \\
          \p_g \scl = 0:  \qquad & c_g g_H = \gamma. \notag
          \end{align}
\end{subequations}
To obtain the explicit form of the dual functional in terms of the dual fields alone, it is convenient to write the terms in $U$ in the Lagrangian
\begin{equation}\label{eq:primal_Lagrangian}
    \begin{aligned}
        \scl(U,\dee, \bar{U}) & = v(- \lambda_x - r \mu_t) + \half c_v (v - \bar{v})^2 + g (- \gamma) + \half c_g g^2 + s(- \gamma_x) - \rho s^2 + \half c_s s^2\\
        & \quad + e(\lambda_t + E_l \mu_x - \beta_t) + E_* e_*^2 \mu_x - \mu_x E_* (e - e_*)^2 + \half c_e (e - \bar{e})^2 \\
        & \quad + X \left(\mu_x + \half \rho d \right) + \half c_X X^2 + d \left( - \beta + \half \rho X \right) + \half c_d \left(d - \bar{d}\right)^2
    \end{aligned}
\end{equation}
in terms of $U - \bar{U}$ through the substitution $U = (U - \bar{U}) + \bar{U}$:
\begin{subequations}\label{eq:Lagrangian}
    \allowdisplaybreaks
    \begin{align}
        \scl(U,\dee, \bar{U}) & = (v - \bar{v}) (- \lambda_x - r \mu_t) + \half c_v (v - \bar{v})^2  + \bar{v} (- \lambda_x - r \mu_t) \notag\\
        & \quad + g (- \gamma) + \half c_g g^2 + s(- \gamma_x) - \rho s^2 + \half c_s s^2 \notag\\
        & \quad + (e - \bar{e}) (\lambda_t + E_l \mu_x - \beta_t) + E_* e_*^2 \mu_x + \bar{e} (\lambda_t + E_l \mu_x - \beta_t) \notag\\
        & \quad - \mu_x E_* (e - \bar{e})^2 - 2 \mu_x E_* (e - \bar{e})(\bar{e} - e_*) - \mu_x E_* (\bar{e} - e_*)^2 \notag\\
        & \quad X(\mu_x + \half \rho (d - \bar{d} + \bar{d}) + \half c_X X^2 \notag\\
        & \quad + (d - \bar{d} + \bar{d}) \left(\half \rho X - \beta \right) + \half c_d \left(d - \bar{d}\right)^2 \notag\\
        & = \notag \\
        & \quad (v - \bar{v}) (- \lambda_x - r \mu_t) \quad + \quad \half c_v (v - \bar{v})^2 \notag\\
        & \quad + g (- \gamma) \quad + \quad \half c_g g^2 \notag \\
        & \quad + s(- \gamma_x) \quad + \quad \half (c_s - 2 \rho) s^2 \notag \\
        & \quad + (e - \bar{e}) \big( \lambda_t + E_l \mu_x - 2 E_* \mu_x (\bar{e} - e_*) - \beta_t \big) \quad + \quad \half (c_e - 2 E_* \mu_x) (e - \bar{e})^2 \tag{\ref{eq:Lagrangian}} \\
        & \quad + (d - \bar{d}) (- \beta) \quad + \quad X (\mu_x + \rho \bar{d}\,) \notag \\
        & \quad + \half \big( c_d (d - \bar{d}) + \rho X \big) (d - \bar{d}\,) \quad + \quad \half \big(\rho (d - \bar{d}) + c_X X \big) X \notag\\
        & \quad + \bar{v} (- \lambda_x - r \mu_t) \ + \ \bar{e} (\lambda_t \ + \ E_l \mu_x - \beta_t) \ + \ E_* \mu_x \big( 2 \bar{e} e_* - \bar{e}^2 \big) \ - \ \beta \bar{d}. \notag
    \end{align}
\end{subequations}
Making the substitution $U \to U_H$ from the DtP mapping \eqref{eq:dtp} into the expression for the Lagrangian $\scl$ in \eqref{eq:Lagrangian} with the notation
\[
\calJ(\dee, \bar{U}) : = \begin{bmatrix}
    - \beta \\
    \mu_x + \rho \bar{d}
\end{bmatrix},
\]
we obtain
\begin{subequations}\label{eq:dual_Lagrangian}
    \allowdisplaybreaks
    \begin{align}      
        \scl(U_H(\dee, \bar{U}),\dee, \bar{U}) & = \left( -1 + \half \right) \left( \frac{1}{c_v} (\lambda_x + r \mu_t)^2  + \frac{1}{c_g} \gamma^2 + \frac{1}{(c_s - 2 \rho)} \gamma_x^2 \right)\notag\\
        & \quad + \left( -1 + \half \right) \frac{1}{(c_e - 2 E_* \mu_x)}\Big( - \lambda_t - E_l \mu_x + 2 E_* \mu_x (\bar{e} - e_*) + \beta_t \Big)^2 \notag\\
        & \quad  - \calJ \cdot \mathbb{K}^{-1} \calJ + \half \left( - \mathbb{K}^{-1} \calJ \right) \cdot \mathbb{K} \left( - \mathbb{K}^{-1} \calJ \right)  \notag\\
        & \quad + \bar{v} (- \lambda_x - r \mu_t) \ + \ \bar{e} (\lambda_t \ + \ E_l \mu_x - \beta_t) \ + \ E_* \mu_x \big( 2 \bar{e} e_* - \bar{e}^2 \big) \ - \ \beta \bar{d}. \notag
    \end{align}
\end{subequations}
Then the \emph{dual functional}, $S[D]$, is defined, in terms of the \emph{dual Lagrangian},
\begin{subequations}\label{eq:dual_L}
    \allowdisplaybreaks
    \begin{align} 
     \scl(U_H(\dee, \bar{U}),\dee, \bar{U}) & = - \half \bigg[ \frac{1}{c_v} (\lambda_x + r \mu_t)^2  + \frac{1}{c_g} \gamma^2 + \frac{1}{(c_s - 2 \rho)} \gamma_x^2 \notag\\
    & \qquad \qquad + \frac{1}{(c_e - 2 E_* \mu_x)}\Big( -\lambda_t - E_l \mu_x + 2 E_* \mu_x (\bar{e} - e_*) + \beta_t \Big)^2 + \calJ \cdot \mathbb{K}|_\rho^{-1} \calJ \bigg] \notag \\
    & \qquad + \bar{v} (- \lambda_x - r \mu_t) \ + \ \bar{e} (\lambda_t \ + \ E_l \mu_x - \beta_t) \ + \ E_* \mu_x \big( 2 \bar{e} e_* - \bar{e}^2 \big) \ - \ \beta \bar{d}, \notag
    \end{align}
\end{subequations}
by
\begin{subequations}\label{eq:dual_S}
    \allowdisplaybreaks
    \begin{align} 
    S[D] & := \int_0^{t_f} \int_0^L \scl(U_H(\dee, \bar{U}),\dee, \bar{U}) \, dx dt \notag\\
 & \quad - \int_0^{t_f} v^l(t) \lambda(0,t) \, dt + \int_0^L e^0(x) \lambda(x,0) \, dx - \int_0^L r v^0(x) \mu(x,0) \, dx - \int_0^{t_f} f(t) \mu(1,t) \, dt. \tag{\ref{eq:dual_S}}
    \end{align}
\end{subequations}
\underline{Remark}: Recall that extremals of the dual functional correspond to the system \eqref{eq:primal} and side conditions \eqref{eq:ibc} with the replacement $U \to U_H$. Moreover, for $\bar{U}$ a solution of (\ref{eq:primal}-\ref{eq:ibc}) it is clear that $D = 0$ is a critical point of $S[D,\bar{U}]$ since the DtP mapping \eqref{eq:dtp} in this case gives $U_H = \bar{U}$. Thus, for each solution of the primal PDE problem (\ref{eq:primal}-\ref{eq:ibc}), there exists at least one dual variational principle whose extremal yields, through the DtP mapping, the solution of the primal problem in question. This is a consistency check for our dual formulation.

Having seen the above fundamental consistency between the dual variational problem and the primal PDE system, we now consider a closely related functional with better properties for analysis and approximation given by
\begin{equation*}
    \begin{aligned}
        \tilde{S}[D] &:= \inf_U \widehat{S}[U,D]  = \int_0^{t_f} \int_0^L \inf_U \scl(U, \dee, \bar{U}) \, dx dt\\
 & \quad - \int_0^{t_f} v^l(t) \lambda(0,t) \, dt + \int_0^L e^0(x) \lambda(x,0) \, dx - \int_0^L r v^0(x) \mu(x,0) \, dx - \int_0^{t_f} f(t) \mu(1,t) \, dt. 
    \end{aligned}
\end{equation*}

We now explicitly work out the Lagrangian of $\tilde{S}$. Denoting 
\[
\begin{bmatrix} d - \bar{d} & X \end{bmatrix}^T =: P(U,\bar{U})
\]
and noting that
\begin{equation*}
\begin{aligned}
    (d - \bar{d}) (- \beta) \quad + \quad X (\mu_x + \rho \bar{d}\,) & = P \cdot \calJ\\
    \big( c_d (d - \bar{d}) + \rho X \big) (d - \bar{d}\,) \quad + \quad  \big(\rho (d - \bar{d}) + c_X X \big) X & = P \cdot \mathbb{K} \,P,
\end{aligned}
\end{equation*}
\begin{equation}\label{eq:tildeL}
\begin{aligned}
     \tilde{\scl}(\dee, \bar{U})  := \inf_U \scl(U,\dee,\bar{U}) & = \begin{cases}
         \scl(U_H(\dee,\bar{U}),\dee, \bar{U}) \quad  \mbox{ if all the following conditions are met: } \\
                  (c_s - 2 \rho) \geq 0 \, ; \quad (c_e - 2 E_* \mu_x) \geq 0 \, ; \quad \mathbb{K} \mbox{ is positive-semidefinite} \,;\\
                  \gamma_x = 0 \mbox{ if } (c_s - 2 \rho) = 0 \,; \\
                  \lambda_t + E_l \mu_x - 2 E_* \mu_x (\bar{e} - e_*) - \beta_t = 0 \mbox{ if } (c_e - 2 E_* \mu_x) = 0 \, ;\\
                  \calJ(\dee, \bar{U}) \mbox{ is orthogonal to eigenspace of } \mathbb{K}(\dee) \mbox{ with 0 eigenvalue}. \\
                   \\
     - \infty \quad \mbox{ otherwise}.
     \end{cases}\\
     & \qquad \mbox{(The definition of the mapping } U_H \mbox{ for all cases is discussed below).}
\end{aligned}
\end{equation}
This is so as follows: in this case of at most quadratic nonlinearity in the primal equations with a quadratic $H$, the Hessian $\p_{UU} \scl(\cdot, \dee, \bar{U})$ does not vary with $U$. When the strict positiveness conditions are met, the Hessian is positive-definite and $\scl(\cdot,\dee,\bar{U})$ has a unique minimum given by the value stated. When the conditions are met with at least one equality, i.e., the Hessian is positive semi-definite but not positive-definite, the minimizer is not unique, but the minimum is unique and given as listed, with $U_H$ interpreted as the unique orthogonal projection of any such minimizer on the orthogonal subspace of the null-space of the Hessian. 

Indeed, denoting the coefficient of the linear terms in $(U - \bar{U})$ in $\scl$ as $\mathcal{P}$, $\scl = (U - \bar{U}) \cdot (\p_{UU} \scl) \ (U - \bar{U}) + \mathcal{P} \cdot (U - \bar{U}) \, + $ terms independent of $U$, and writing the (symmetric) Hessian in spectral decomposition, $\p_{UU} \scl = \sum_i a_i e_i \otimes e_i$, $(a_i, e_i)$ being eigenpairs, the $\inf_U \scl$ is attained when the condition $\sum_i \left( a_i (U - \bar{U}) \cdot e_i  + \mathcal{P} \cdot e_i \right) e_i = 0$ is satisfied and the minimum value of $\scl$ is given by $\sum_i a_i \left( a_i (U - \bar{U}) \cdot e_i \right)^2 + (\mathcal{P} \cdot e_i) ((U - \bar{U}) \cdot e_i) \, + $ terms independent of $U$. It is now clear that if $\mathcal{P}$ were not orthogonal to the `null eigenspace', the eigenspace of the Hessian with null eigenvalue, for which the quadratic terms vanish (possibly only one) in the last expression, $U - \bar{U}$ could be chosen to have a suitable component along the null eigenspace and the value of $\scl$ could be made unbounded below. Of course, in that case the linear algebraic system, $\p_{UU} \scl \ (U - \bar{U}) = - \mathcal{P}$ would not have a solution as well, equivalently seen from $\sum_i \left( a_i (U - \bar{U}) \cdot e_i  + \mathcal{P} \cdot e_i \right) e_i = 0$. Finally, when $\mathcal{P}$ is orthogonal to the null eigenspace of the Hessian, and we do not have strict positive definiteness, it is also clear from the decomposition of the problem into spectral form that the solution for $U - \bar{U}$ is unique only up to addition of components from the null-space of the Hessian, with the minimum value of $\scl$ unaffected. This non-uniqueness cannot be allowed in the DtP mapping as it has to be a function of the dual variables (and a one-to-many relation is not allowed). Consequently, the definition of $U_H$ is augmented as stated above, without any loss of generality.

There are two noteworthy features of the definitions above: 
\begin{itemize}
    \item First, since $\scl(U,\dee, \bar{U})$ is necessarily affine in $\dee$ and therefore concave in it, $\tilde{\scl}(\dee, \bar{U})$ is concave in $\dee$ (but not necessarily strictly concave) by a theorem of convex analysis.
    \item Second, since the functional $\widehat{S}[U,D]$ is necessarily affine and hence continuous (in an appropriate weak sense) in the dual fields $D$, the functional $\tilde{S}[D]$ is upper semicontinuous (again, by a theorem of convex analysis). Assuming the space of dual fields is a closed convex set (a natural assumption), then the existence of a maximizer of $\tilde{S}$ depends only on the coercivity of $-\tilde{S}$.
\end{itemize}

From a more practical point of view, in looking for a maximizer of $\tilde{S}[D]$ one might as well restrict attention to dual fields for which $\tilde{\scl}(\dee,\bar{U}) = \scl(U_H(\dee, \bar{U}), \dee, \bar{U})$ almost everywhere in $(0,L) \times (0,t_f)$, the conditions on such dual fields being specified in \eqref{eq:tildeL}. Since for such fields the functional $\tilde{S}[D]$ and $S[D]$ coincide, and a maximizer of a functional is also a critical point/extremal (at least formally), solving the \emph{maximum} problem
\begin{equation}\label{eq:sup_inf}
    \argmax_D \tilde{S}[D]
\end{equation}
generates a dual solution which, through the DtP mapping, produces a solution to our primal PDE system \eqref{eq:primal}-\eqref{eq:ibc}.

The advantage of the formulation \eqref{eq:sup_inf} is that by restricting the search-space of dual solutions as described in the previous paragraph, we essentially work with a \emph{concave maximization} problem which has a single maxima, regardless of the convexity/monotonicity properties of the primal problem. By using the base state, $\bar{U}$, judiciously, one can home in on different primal solutions in a reliable manner (cf.~\cite{KA1, KA2}).
\section{Concluding remarks}\label{sec:concl}
A procedure to treat the Second Law as an inequality to be solved on par with the other balance laws in continuum thermomechanics, instead of as a restriction on constitutive assumptions, has been discussed. The approach allows for an accommodation of the approximate nature of constitutive assumptions as well.

We close with the following observations:
\begin{enumerate}

    \item Since our scheme guarantees by construction that if a dual solution to the critical point problem exists then a corresponding primal solution must (at the level of formality assumed in these discussions), it is implied that if no solutions exist to the primal system then neither can a dual extremal exist for such a problem. Thus, unlike the method of Least Squares (the primal PDE system is `squared' to produce an objective to be minimized), the dual scheme cannot produce spurious solutions to the primal problem when looking at first-order optimality conditions.
    \item The excess quantities in our formalism are determined by the assumed constitutive equations, the balance (in)equalities, and boundary and initial conditions. As a result, these contributions to the stress and energy density are nonlocal, in general (a feature much like that of the pressure field for an incompressible material).
    \item The inclusion of prescribed base states (in space-time) within our solution scheme allows for the seamless accommodation of approximate information on specific motions from experiments to be included in the modeling approach.
    \item Our formulation of the dissipation introduces some intriguing flexibility in problem solving that needs to be further explored. For example, instead of minimizing  $-\half |c_s| s^2, c_s < 0$, one could seek to maximize $\frac{1}{(1-p)} |c_s| |s|^{1 - p}, c_s < 0, 0\leq p < 1$ (see Remark 5. below), and this is only one of many options.
    \item In \cite[Sec.~7.1]{needleman2023discrete} a dynamic viscoplastic model of a bar is considered where the constitutive equation is intentionally made to violate the Clausius-Duhem Inequality (we leave aside for the moment the philosophical question of whether an explicitly time-dependent parameter ($\dot{a}$ in that work, called $A$ below) can appear in material response and whether that is consistent with the admissible processes allowed in the Coleman-Gurtin formalism \cite{coleman1967thermodynamics} of thermodynamics with internal variables). Within our setting, this can be modeled using the assumptions 
    \[
    X = 0, \qquad T = \sigma(e, e^p) = E\,(e - e^p), \quad E = \mbox{ Young's modulus}
    \]
    along with the equations
    \begin{equation}
        \begin{aligned}
            rv_t - \sigma(e,e^p)_x & = 0\\
            e^p_t - A(t) \frac{\sigma}{|\sigma|} p(\sigma) & = 0; \quad p \mbox{ a (power-law, positive) function of } \sigma; \quad t \mapsto A(t) \in \mathbb{R} \mbox{ given}\\
            \sigma(e, e^p) e^p_t - a_t - s^2 & = 0,
        \end{aligned}
    \end{equation}
    with boundary conditions on $v$ and initial conditions on (effectively) $(e,v, e^p)$. In the example of Needleman, $A$ can have both positive and negative signs in different intervals of time.

    Choosing $(v,e, e^p, a, s)$ as the primal variables in our setting, the procedure of Sec.~\ref{sec:dual} can be carried out on this system and solutions obtained, consistent with the Clausius-Duhem Inequality not being violated. It is clear, of course, that for such a solution, the `classical' elastoplastic dissipation 
    \[
    \sigma(e, e^p) e^p_t \mbox{ is not necessarily } \geq 0 \mbox{ for all } (x,t).
    \]
    In \cite{needleman2023discrete} it is said ``A suitable continuum mechanics
    expression of the second law of thermodynamics \ldots that allows the Clausius–Duhem inequality to be violated
    locally (while maintaining overall stability) remains to be developed.''

    The methodology proposed in this work may be considered as one possible start on addressing this question.
    \item The problem above is an example of a situation where the mechanical equations constrain the solution adequately and the Second Law with classical ingredients would be overconstraining. Another extreme is the situation when the mechanical equations are underconstraining and the Second Law is thought of as one device to provide adequate constraints (e.g., Dafermos \cite{dafermos1973entropy, dafermos2012maximal}, Slemrod and co-workers \cite{gimperlein2024least}). This is also the point of view of Abeyaratne and Knowles \cite{abeyaratne2006evolution} in the theory of phase transition in nonlinear elasticity, where the Second Law is used to provide guidance on kinetic laws for the motion of phase boundaries and their nucleation. A somewhat different point of view of conferring adequate constraints to the problem is through capillary-viscosity regularizations arising from the works of Slemrod \cite{slemrod1983admissibility,slemrod1989limiting} and Truskinovsky \cite{truskinovsky1994normal}.

    In the present setting, a problem definition to probe this point of view could be to use the following set of equations (setting $X= 0, a =0$):
    \begin{subequations}\label{eq:elast_PT}
    \allowdisplaybreaks
    \begin{align}
        v_x - e_t & = 0 \notag\\
        r v_t - (\hat{T}(e))_x  & = 0 \notag\\
        e_t - d & = 0 \tag{\ref{eq:elast_PT}}\\
        \hat{T}(e)d - (F(e))_t - s^2 & = 0 \notag\\
        s_x - g & = 0, \notag
     \end{align}
    \end{subequations}
    with
    \begin{subequations}\label{eq:PT_H}
    \begin{align}
        & \quad H(U, \bar{U}) := \left( \frac{c_s}{(1 - p)} |s|^{(1 - p)} + \half  c_g g^2 \right) + \half c_d \left(d - \bar{d} \right)^2 + \half c_e \left( e - \bar{e} \right)^m + \half c_v \left( v - \bar{v} \right)^2, \notag\\
         &  c_g, c_d, c_e, c_v  > 0, \quad  c_s < 0, \quad 0\leq p < 1, \quad  \mbox{ with } H \mbox{ dimensionally consistent,} \notag\\
         & m \mbox{ chosen such that } \hat{T}(e) + \half c_e \left( e - \bar{e} \right)^m \mbox{ is monotone increasing in } e \mbox{ (e.g., see \cite{singh2024hidden})}; \notag\\
         & \mbox{the choice of the dependence of } H \mbox{ on } s \mbox{ is motivated by the form of the resulting } \notag\\
         & \mbox{DtP mapping in the absence of the field } \gamma. \notag
    \end{align}
    \end{subequations}
    An alternative formulation with fewer variables is
       \begin{subequations}\label{eq:elast_PT_2}
    \allowdisplaybreaks
    \begin{align}
        v_x - e_t & = 0 \notag\\
        r v_t - (\hat{T}(e))_x  & = 0 \notag\\
        \left(\hat{T}(e)v \right)_x -\left(\half rv^2\right)_t - (F(e))_t - s^2 & = 0 \notag\\
        s_x - g & = 0, \notag\\
        \mbox{with } c_d = 0, \quad d & \equiv 0.\notag
     \end{align}
    \end{subequations}
    First, we consider the case where $g = 0, \gamma = 0$ (recall that $\gamma$ is the dual field corresponding to the last equation in \eqref{eq:elast_PT}). Forming a Lagrangian and a dual action for the system \eqref{eq:elast_PT}-\eqref{eq:PT_H} and proceeding as in Sec.~\ref{sec:dual}, we note that the first variation of the dual action (set to 0) is the primal PDE problem stated in weak form (in the sense of distributions). This implies that the Euler-Lagrange equations and side conditions \emph{include the jump conditions} corresponding to the balance (in)equalities. Then, it is possible to have solutions where the dissipation is concentrated on space-time curves of discontinuity, as is the field $s$. Indeed, by the Remark noted in Sec.~\ref{sec:dual}, if one chooses a set of $(\bar{v},\bar{e},\bar{d})$ fields that represent a shock solution of classical elastodynamics with a shock(s) and positive dissipation on these space-time shock curves, then the dual scheme will also admit that triple as a DtP mapping generated primal solution. Now, if the triple $(\bar{v},\bar{e},\bar{d})$ is close to such a shock solution, it may be expected that the dual scheme will recover the latter solution. This is because of the local degenerate ellipticity of the dual scheme \cite{ach_HCC}, as demonstrated in \cite{KA2} in the context of the inviscid Burgers equation with base states that are, piecewise in time, solution to Burgers equation with small viscosity. Thus the effect of the base states $(\bar{v}, \bar{e}, \bar{d})$ may be viewed as akin to a selection criterion in the model (cf.~\cite{abeyaratne2006evolution,slemrod1983admissibility,truskinovsky1994normal}).

    With $g, \gamma$ as active fields in the model, the dissipation may even not be allowed to concentrate on curves, thus bringing a regularizing effect assuming, of course, that there exists smooth solutions to classical elastodynamics close to shock solutions. If indeed this bears out, it becomes a matter of modeling taste whether, in the absence of physical knowledge of appropriate higher-order theories, whether this minimal postulate of smooth dissipation is a better modeling option or not - certainly, it does not result in higher order equations. As some evidence that these speculations may have a basis, in \cite{singh2024hidden} an elastodynamic equilibrium for a bar with a double-well elastic energy density (and negative stiffness in intervals of non-zero length) is computed by the dual scheme in a problem where the primal Cauchy problem lacks continuous dependence w.r.t initial data.

    We note here that the equations to be solved by the dual scheme (e.g., \eqref{eq:elast_PT}) do not involve any regularizations or extra parameters, even though the auxiliary function $H$ does.

    \item It is interesting to note that jumps (strong discontinuities) in the primal fields can be represented by the dual scheme under appropriate circumstances, even though the dual problem is a second-order boundary value problem in space-time domains. As shown in the contexts of the heat and linear transport equations in \cite{KA1}, the dual problem is non-elliptic in these situations (also see \cite{KA2}) but arises from a concave dual variational principle. Thus gradient (weak) discontinuities in dual fields are admitted which, through the DtP mapping, produce discontinuous primal fields when needed, with curves of dual weak discontinuities lining up exactly with characteristic curves of the primal problem.
    \item The constitutive equation for stress \eqref{eq:stress_const} comes from a cubic energy density that is neither bounded above or below. Thus, there can be no hope of defining energy minimizing solutions of the bar utilizing such an energy density, in general. However, solutions to the Euler-Lagrange equations of an energy functional with such a density can definitely exist (cf.~\cite{acharya2024variational}). The proposed dual scheme sets up a variational problem for obtaining weak solutions to these E-L equations, amenable to approximation, and potential analysis by the Direct method of the Calculus of Variations, both in the quasistatic and dynamic cases.
    \item We note that, augmenting the definition of an admissible thermodynamic process in \cite{coleman1967thermodynamics} to include the excess fields discussed herein, any thermodynamic process that our proposed procedure produces would be compatible with the Second Law, which is the essential requirement of Coleman and Noll \cite[(4.4)]{CN} and Coleman and Gurtin \cite[Sec.~5]{coleman1967thermodynamics}.
\end{enumerate}
\printbibliography
\end{document}